\documentclass[12pt]{article}
\usepackage{amsmath,amssymb}
\usepackage{epsfig}
\usepackage{amsfonts}
\usepackage{amsmath,amssymb}
\usepackage{epsfig}
\def\dref#1{(\ref{#1})}

\begin{document}

\vspace*{1\baselineskip} \begin{center}\Large\bf Are networks with
more edges easier to synchronize? \footnote {\small
This work is jointly supported by the National Natural Science
Foundation of China under grant 60674093, the Key Projects of
Educational Ministry under grant 107110 and the  City University  of
Hong  Kong  under  the  Research  Enhancement Scheme  and  SRG
grant  7002134.}
\end{center} \vspace*{1\baselineskip}

\centerline {Zhisheng Duan$^1$,  ~Wenxu Wang$^2$, ~ Chao Liu$^1$,
~and ~Guanrong Chen$^{1,2}$ }

\vspace*{0.5\baselineskip}
\begin{center}
$^1$State Key Laboratory for Turbulence and Complex Systems,
Department of Mechanics and Aerospace Engineering, College of
Engineering, Peking University, Beijing 100871, P. R. China\\
$^2$ Department of electronic Engineering, City University of Hong
Kong, Hong Kong
 \\ Email: duanzs@pku.edu.cn,  {\it Fax}: (8610)62765037
\end{center}
\vskip 0.5cm {\it Abstract.} \,\, In this paper, the relationship
between the network synchronizability and the edge distribution of
its associated graph is investigated. First, it is shown that adding
one edge to a cycle definitely decreases the network
sychronizability. Then, since sometimes the synchronizability can be
enhanced by changing the network structure, the question of whether
the networks with more edges are easier to synchronize is addressed.
It is shown by examples that the answer is negative. This reveals
that generally there are redundant edges in a network, which not
only make no contributions to synchronization but actually may
reduce the synchronizability. Moreover, an example shows that the
node betweenness centrality is not always a good indicator for the
network synchronizability. Finally, some more examples are presented
to illustrate how the network synchronizability varies following the
addition of edges, where all the examples show that the network
synchronizability globally increases but locally fluctuates as the
number of added edges increases.



\section{Introduction and problem formulation}
 Systems composing of dynamical units are ubiquitous in nature,
ranging from physical to technological, and to biological fields.
These systems can be naturally described by networks with nodes
representing the dynamical units and links representing the
interactions among them. The topologies of such networks have been
extensively studied and some common architectures have been
discovered \cite{Review1,Review2}. The small-world property, for
example, characterized by short average distance and high clustering
among nodes, is one of the most common properties shared by many
real networks \cite{wat98}. More significantly, many networks show
high heterogeneity of node connectivity, which typically possesses a
power-law distribution, named scale-free networks \cite{bar99}. It
is known that these topological characteristics have strong
influence on the dynamics of the structured systems, such as
epidemic spreading, traffic congestion, collective synchronization,
and so on \cite{Review3,Review4}. From this viewpoint,
systematically understanding the network structural effects on their
dynamical processes is of both theoretical and practical importance.

Synchronization behavior, in particular, as a widely observed
phenomenon in networked systems has received a great deal of
attention in the past decades
\cite{bar02,luw04,bel05,Mot05PRE,Cha05,Zhou061,atay06,wang06,sor07,ste07,lu04,wang02}.
Oscillator network models have been commonly used to characterize
synchronization behaviors. In this setting, a synchonizability
theorem provided by Pecora and Carroll \cite{pec98} indicates that
the collective synchronous behavior of a network is completely
determined by the network structure, independent of individual
oscillator dynamics, provided that the network coupling strength
satisfies some strong conditions. In this framework, it has been
found that, compared to regular lattices,  small-world networks have
remarkably better synchronizability \cite{syncSW}. In contrast to
small-world networks, scale-free networks tend to inhibit
synchronization, although they have much shorter average distances
than small-world networks \cite{nis03} which are generally deemed to
be advantageous  for synchronization. Therefore the node betweenness
centrality was provided as a good indicator to the network
synchronizability \cite{hong04}. Since the synchronizability is
correlated with  many topological properties, a natural question is
which property is the most significant to the synchronizability?
Donetti et al. \cite{don05} gave a good answer to this question by
an optimization argument. They pointed out that a network with
optimized synchronizability should have an extremely homogeneous
structure, i.e., the distributions of some fundamental  topological
properties should be very narrow. Their work provides a constructive
approach to the issue of networked synchronization, making a big
progress in this area. However, some issues still remain unclear,
e.g., what is the most important topological property for the
synchronizability? And what is the effect of the connectivity
density on the synchronizability? In particular, as admitted by the
authors, this approach  cannot theoretically guarantee to find the
optimal solution.

Motivated by the above works, this paper focuses on the
relationships between the network synchronizability and the edge
distribution of the associated graph. The effects of the connection
patterns of graphs on the synchronizability are analyzed both
theoretically and numerically. It is found that adding an edge to a
cycle of size $N\geq 5$ definitely decreases the network
synchronizability, but the synchronizability may be improved by
changing the cyclic
 structure. However, a further example shows that,
by arbitrarily optimizing the network structures, networks with more
edges are not necessarily easier to synchronize. This implies that
there are redundant edges in the network with respect to
synchronization.

Consider a dynamical network consisting of $N$ coupled identical
nodes, with each node being an $n$-dimensional dynamical system,
described by
\begin{equation}\label{n1}
\dot{x}_i=f(x_i)-c\sum_{j=1}^Na_{ij}H(x_j),\;i=1,2,\cdots,N,
\end{equation}
where $x_i=(x_{i1},x_{i2},\cdots,x_{in})\in \mathbb{R}^n$ is the
state vector of node $i$, $f(\cdot):\mathbb{R}^n\rightarrow
\mathbb{R}^n$ is a smooth vector-valued function, constant $c>0$
represents the coupling strength, $H(\cdot):\mathbb{R}^n\rightarrow
\mathbb{R}^n$ is called the inner linking function, and
$A=(a_{ij})_{N\times N}$ is called the outer coupling matrix or
topological  matrix, which represents the coupling configuration of
the entire network. This paper only considers the case that  the
network is diffusively connected, i.e., $A$ is irreducible and its
entries satisfy
$$
a_{ii}=-\sum_{j=1,j\neq i}^Na_{ij},\;i=1,2,\cdots,N.
$$
Further, suppose that, if there is an edge between node $i$ and node
$j$, then $a_{ij}=a_{ji}=-1$, i.e., $A$ is a Laplacian matrix. In
this setting, $0$ is an eigenvalue of $A$ with multiplicity 1, and
all the other eigenvalues of $A$ are strictly positive, which are
denoted by
\begin{equation}\label{f1}
0=\lambda_1<\lambda_2\leq\lambda_3\leq\cdots\leq\lambda_N.
\end{equation}

The dynamical network \dref{n1} is said to achieve (asymptotical)
synchronization if
$x_1(t)\rightarrow x_2(t)\rightarrow\cdots\rightarrow
x_N(t)\rightarrow s(t),\;\textrm{as}\; t\rightarrow \infty,$
where, because of the diffusive coupling configuration,  the
synchronous state $s(t)\in \mathbb{R}^n$ is a solution of an
individual node, i.e., $\dot{s}(t)=f(s(t))$.

It is well known that the eigenratio
$r(A)=\frac{\lambda_2}{\lambda_N}$ of the network structural matrix
$A$ characterizes the synchronizability. The larger the $r(A)$ is,
the better the synchronizability will be. The enhancement of the
network synchronizability and the relationships between $r(A)$ and
the network structural characteristics such as average distance,
node betweenness, degree distribution, clustering coefficient, etc.,
have been well studied \cite{atay05, duan07, mot05, wu05, zhao06,
zhou06}. In particular,  the graph-theoretical method was used to
discuss the network synchronizability in \cite{atay05, duan07}. This
paper further investigates the relationship between the network
edges and its synchronizability by graph-theoretical tools.


Throughout this paper, for any given undirected graph $G$,
eigenvalues of $G$ mean eigenvalues of its corresponding Laplacian
matrix. Notations for graphs and their corresponding Laplacian
matrices are not differentiated, and networks and their
corresponding graphs are not distinguished, unless otherwise
indicated.

\section{Adding one edge to a cycle}

It has been shown by examples that adding edges can either increase
or decrease the network  synchronizability \cite{duan07}, and it was
found \cite{yin06} that in scale-free networks where the nodes are
coupled symmetrically, if some overloaded edges are removed, the
network will become more synchronizable.

 In this section, consider
adding one edge to a given cycle with $N$ $(N\geq 4)$ nodes. In this
case, the added edge definitely decreases the sychronizability. To
show this, the following lemmas are needed.

{\it Lemma 1}\,\cite{mer94, mer98}\, For any given connected graph
$G$ of size $N$, its nonzero eigenvalues indexed as listed in
(\ref{f1}) grow monotonically with the number of added edges; that
is, for any added edge $e$, $\lambda_i(G+e)\geq \lambda_i(G)$, $i=1,
\cdots, N$.

Lemma 1 shows the eigenvalue changes of graphs due to the addition
of edges, but it does not show any information about the eigenratio
$r(A)$. Therefore, this eigenratio  needs to be studied in more
detail.

{\it Lemma 2}\,\cite{wu05,mer94}\, For any given connected graph $G$
of size $N$, its largest eigenvalue $\lambda_N$ satisfies
$\lambda_N\geq d_{max} +1$, with equality if and only if
$d_{max}=N-1$. Further, if $G$ is not a complete graph, then the
smallest nonzero eigenvalue of $G$ satisfies $\lambda_2\leq
d_{min}$. Here $d_{max}$ and $d_{min}$ denote the maximum and
minimum degrees of $G$.

{\it Lemma 3}\,\cite{wu05}\, For any cycle $C_N$ with $N$ ($\geq 4$)
nodes, its eigenvalues are  given by $\mu_1, \cdots, \mu_N$ (not
necessarily ordered as in (\ref{f1})) with $\mu_1=0$ and
$$\mu_{k+1}= 3- \frac {\sin (\frac{3k\pi}{N})}{\sin(\frac{k\pi}{N})}, \,\, k=1,\cdots, N-1.$$


{\it Lemma 4}\,\, Given a connected graph $G$, if the multiplicity
of its smallest nonzero eigenvalue $\lambda_2$ is larger than or
equal to 2, then adding one edge to $G$ can not change this
eigenvalue, i.e., $\lambda_2(G+e)=\lambda_2(G).$

{\it Proof} \,\, This lemma follows from the fact that  $\hbox {rank
} (\lambda_2 I-(G+e))\leq \hbox {rank } (\lambda_2 I-G)+1$. \hfill
$\Box$

By the above lemmas, one can get the following result for cycles.

{\it Theorem 1}\,\, For any cycle $C_N $ with $N\geq 4$ nodes,
adding one edge can not enhance its synchronizability  $r(C_N)$;
specifically, one has $r(C_4+e) = r(C_4)$ and $r(C_N+e) < r(C_N)$
($N\geq 5$).

{\it Proof}\,\, $r(C_4+e) = r(C_4)$ holds obviously. For the case of
$N\geq 5$, by Lemma 2, one has $\lambda_N(C_N+e)>4$. But by Lemma 3,
$\lambda_N(C_N)\leq 4$. And Lemma 3 shows that the multiplicity of
the smallest nonzero eigenvalue $\lambda_2$ of $C_N$ is 2. By Lemma
4, $\lambda_2(C_N+e)=\lambda_2(C_N)$. Therefore, $r(C_N+e) < r(C_N)$
for all $N\geq 5$. \hfill $\Box$

Theorem 1 shows that adding one edge a cycle with  $N\geq 5$ nodes
definitely decreases the network sychronizability, as shown by the
two examples in Figs. 1-5. By simple computation, one obtains that
$r(C_5)=\frac{1.3820}{3.6180}=0.3820$ and
$r(C_5+e\{1,3\})=\frac{1.3820}{4.6180}=0.2993<r(C_5)$;
$r(C_6)=\frac{1}{4}=0.25$,
$r(C_6+e\{1,3\})=\frac{1}{4.4142}=0.2265<r(C_6)$ and
$r(C_6+e\{1,4\})=\frac{1}{5}=0.2<r(C_6)$.

\begin{center}
\unitlength=1cm \qquad \hbox{\hspace*{0.1cm}  \epsfxsize5.5cm
\epsfysize5cm \epsffile{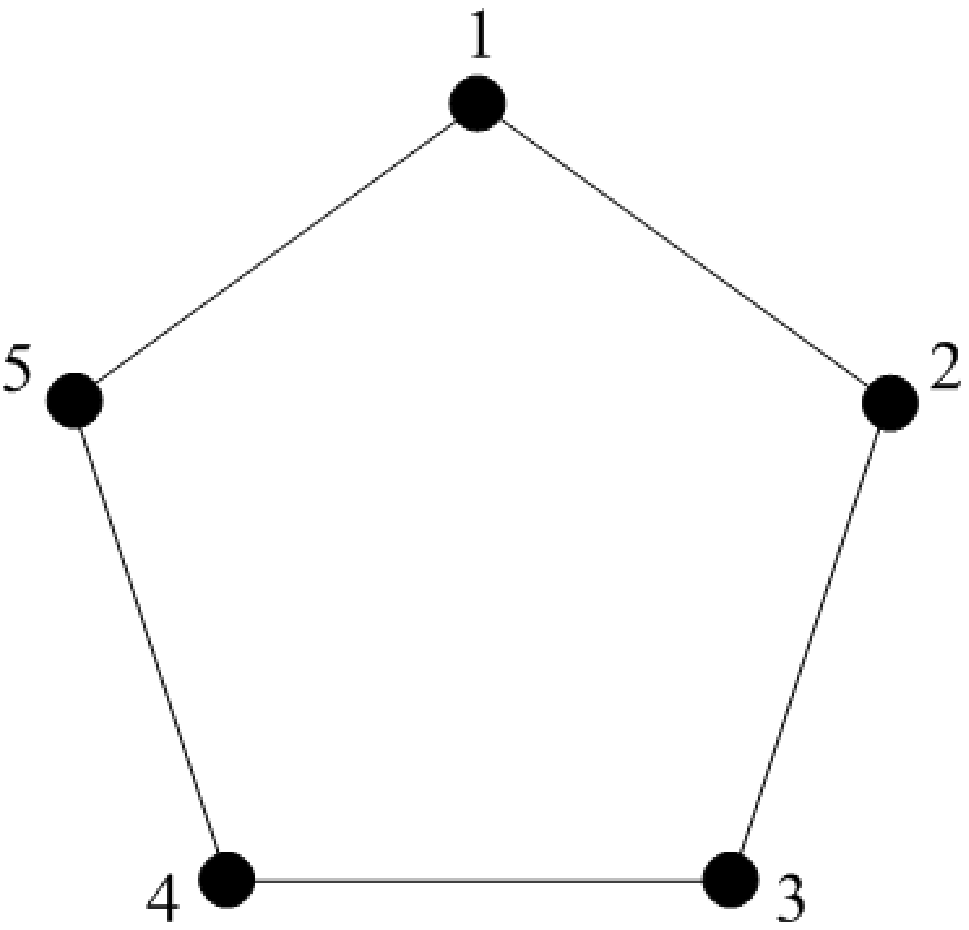}\qquad \epsfxsize5.5cm \epsfysize5cm
\epsffile{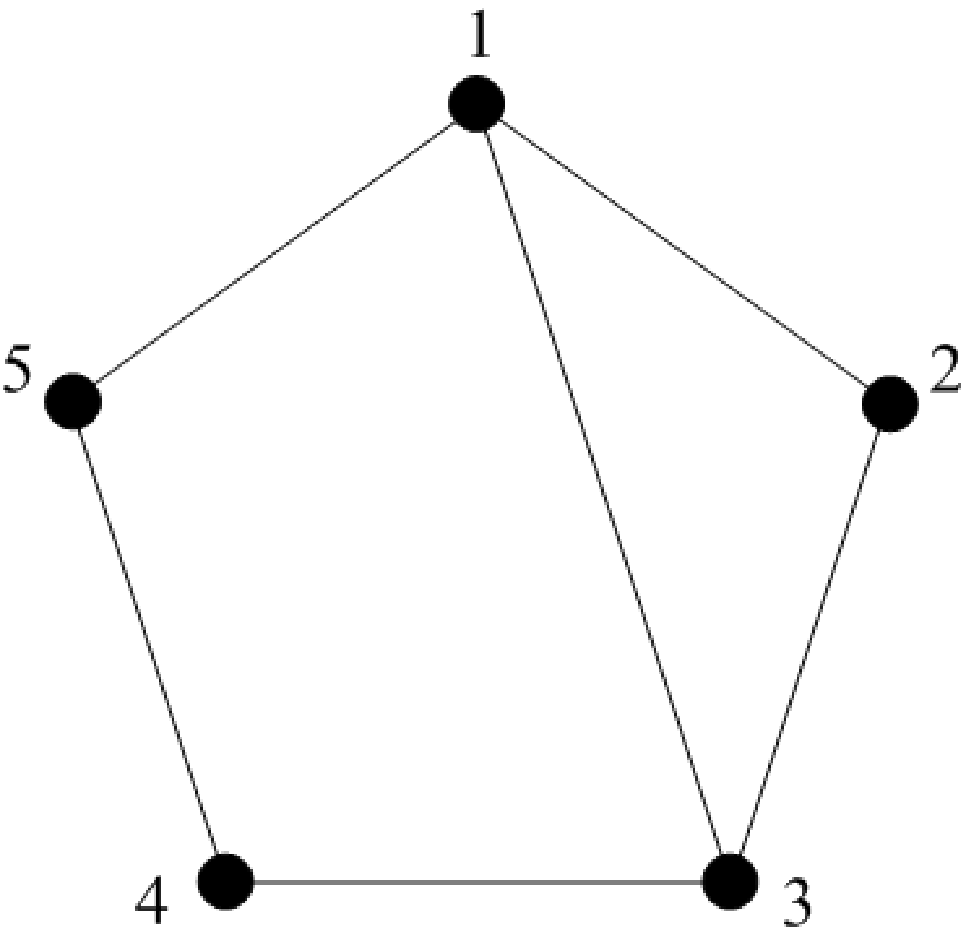} }
\end{center}
\vskip -0.3cm \centerline{\qquad\quad  Fig. 1 \,\, Graph $C_5$
\qquad\qquad\qquad  Fig. 2 \,\, Graph $C_5+e\{1,3\}$}

\begin{center}
\unitlength=1cm \qquad \hbox{\hspace*{0.1cm}  \epsfxsize4.8cm
\epsfysize5cm \epsffile{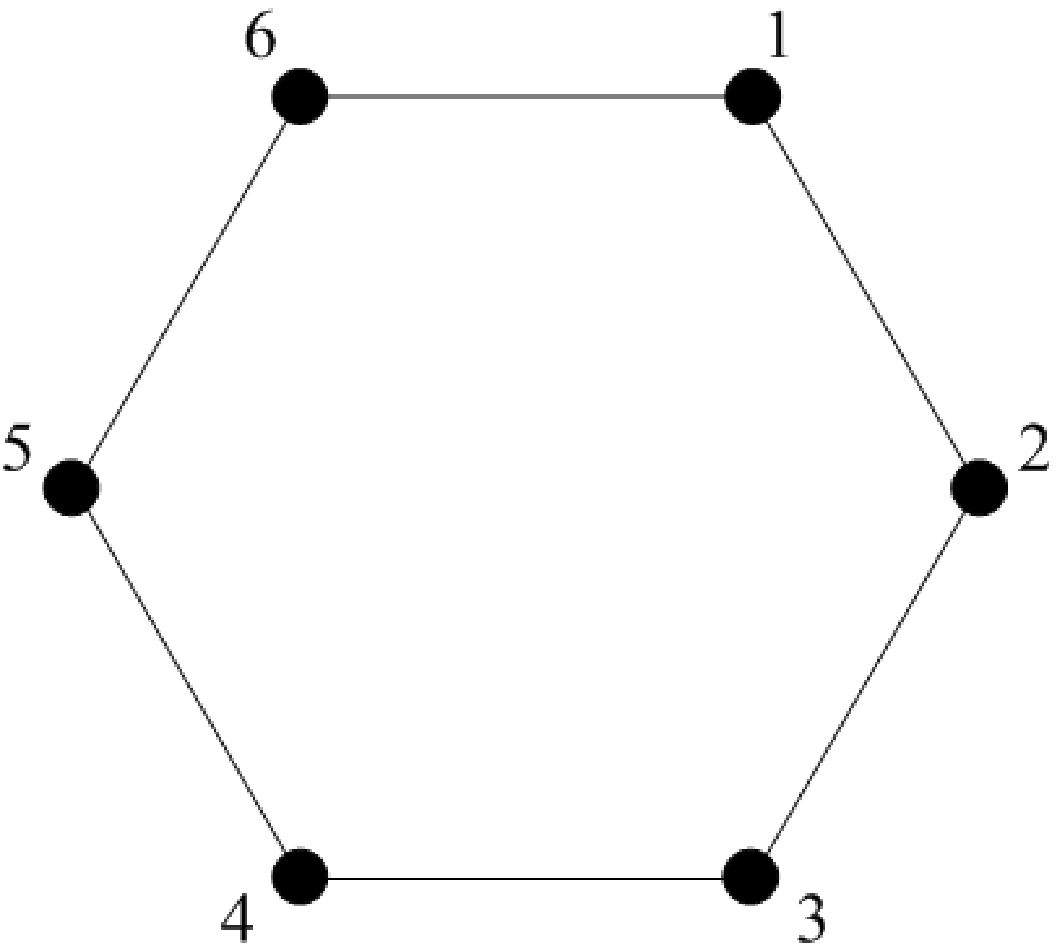}\qquad \epsfxsize4.8cm \epsfysize5cm
\epsffile{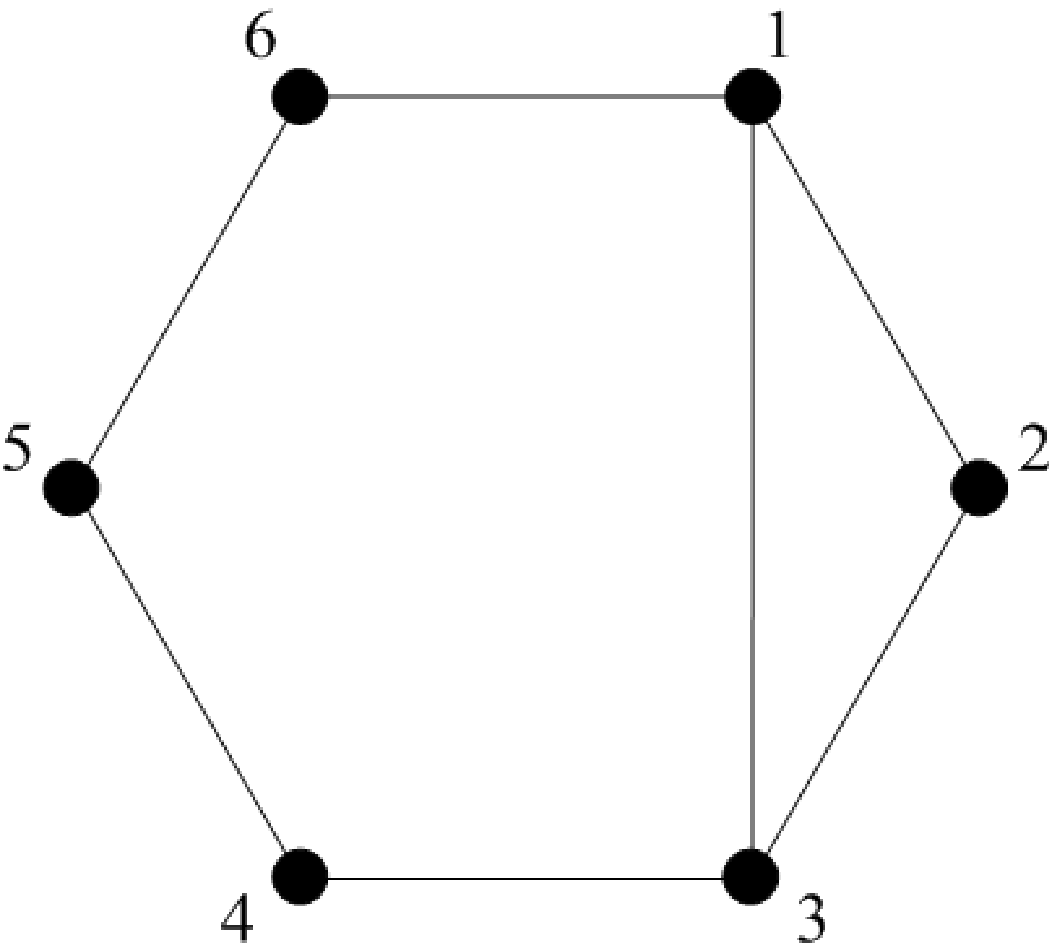} \qquad \epsfxsize4.8cm \epsfysize5cm
\epsffile{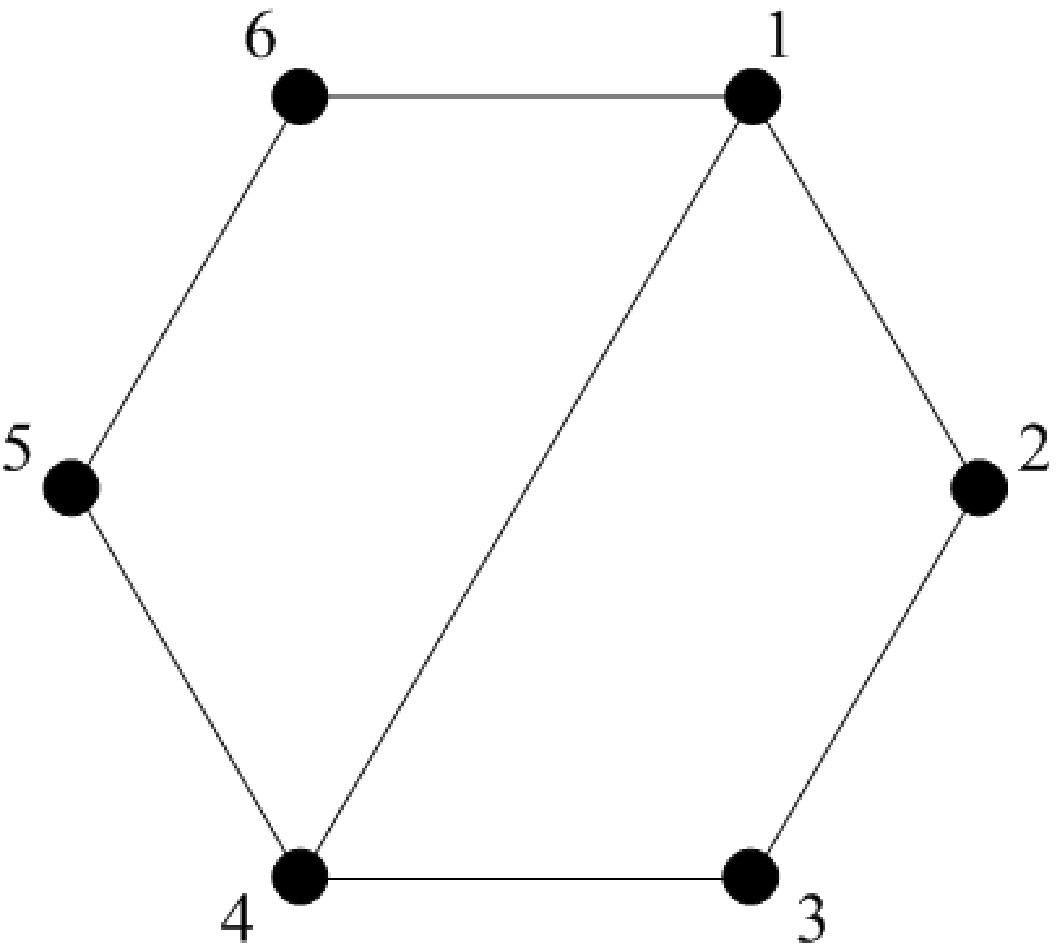} }
\end{center}
\vskip -0.3cm \centerline{  Fig. 3 \,\, Graph $C_6$
\qquad\qquad\qquad Fig. 4 \,\, Graph $C_6+e\{1,3\}$ \qquad  Fig. 5
\,\, Graph $C_6+e\{1,4\}$}

From the above two examples, one can find that the synchronizability
of cycles strictly decreases if only one edge is added, and the
results vary depending upon where the edge is placed, e.g.,
$r(C_6+e\{1,3\})
>r(C_6+e\{1,4\}).$ Considering the optimization of network
structures, $r(C_6+e\{1,3\})$ is still not the best one among all
graphs with 7 edges connecting 6 nodes in a cycle, as demonstrated
in the next section.

\section{Changing the network structure to enhance its synchronizability}

It is shown in the above section that adding one edge to a cycle
decreases its synchronizability. A further question is whether the
synchronizability can be enhanced by changing the network structure
after edge addition. The answer is `yes' in some cases. For example,
one can change $C_5+e\{1,3\}$ to $C_{5o}$ as in Fig. 6, and
$C_6+e\{1,3\}$ to $C_{6o}$ as in Fig. 7. Then,
$r(C_{5o})=\frac{2}{5}=0.4$ and
$r(C_{6o})=\frac{1.2679}{4.7231}=0.2684.$

\begin{center}
\unitlength=1cm \qquad \hbox{\hspace*{0.1cm}  \epsfxsize6cm
\epsfysize5.5cm \epsffile{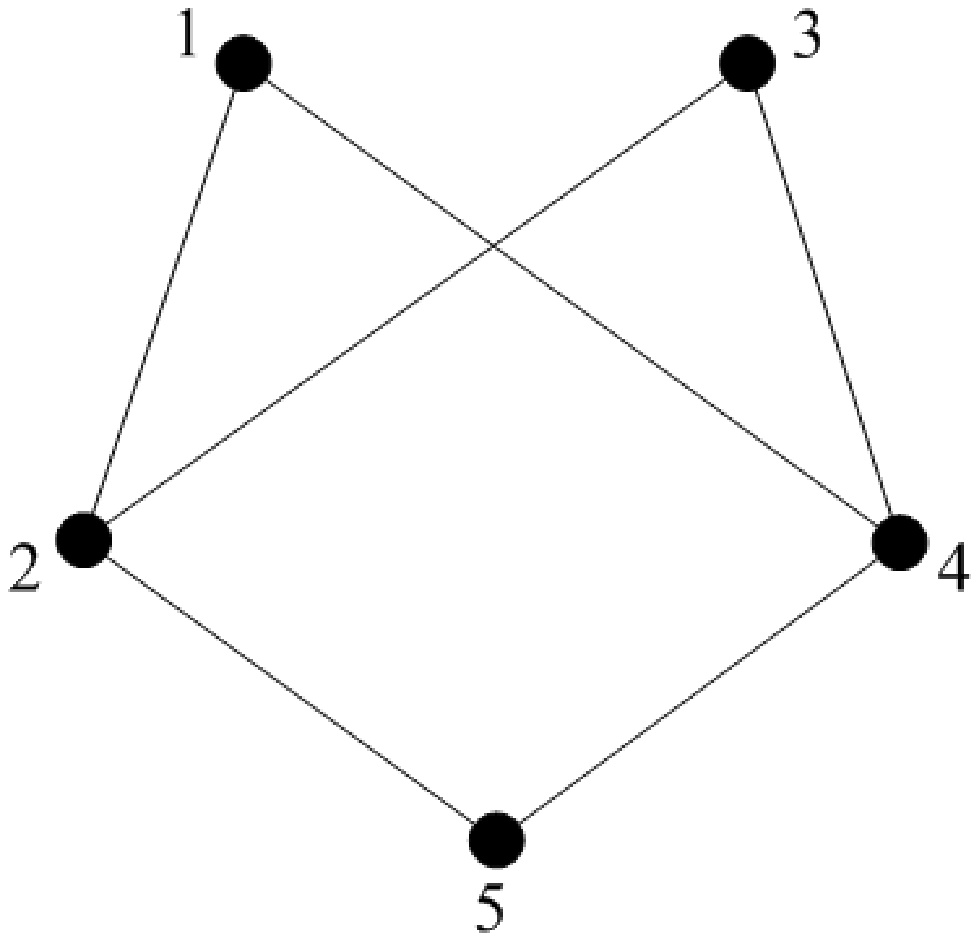}\qquad \epsfxsize5.5cm
\epsfysize5cm \epsffile{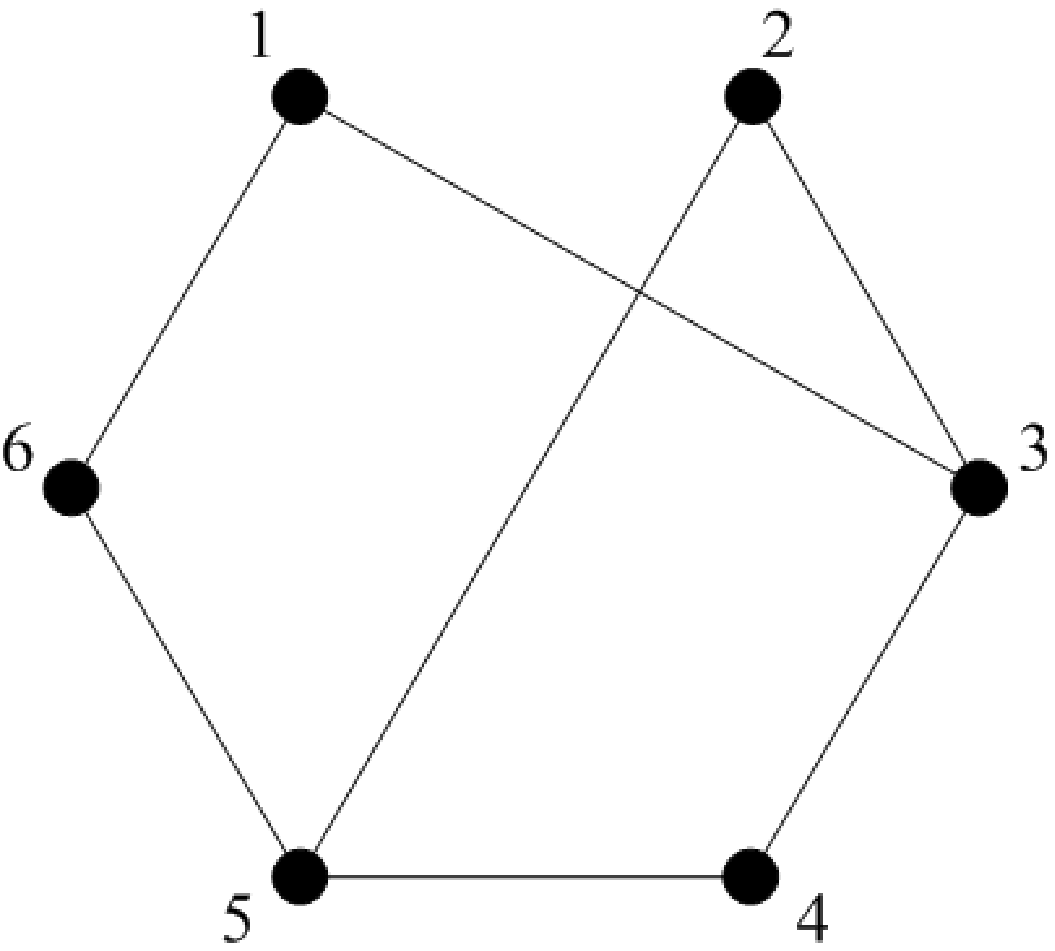} }
\end{center}
\vskip -0.3cm \centerline{\quad  Fig. 6 \,\, Graph $C_{5o}$
\qquad\qquad\qquad \qquad Fig. 7 \,\, Graph $C_{6o}$}

Comparing with the graphs in Figs. 1-5, one can see that  both the
synchronizabilities of $C_{5o}$ and $C_{6o}$ have been improved. In
fact, two cycles share a common edge in Figs. 2, 4 and 5. In this
case, generally the betweenness centrality is large, or the
node-to-node distances are not homogeneous. In comparison, the
network structural characteristics are more homogeneous in Figs. 6
and 7. This is consistent with the result of \cite{don05}. For
simple graphs with a few nodes and edges, as those shown above, one
can compute their eigenvalues to find a good structure for the
synchronizability. However,  for a general graph, how to optimize
the network structure toward the best possible  synchronizability?
Some optimal rules are provided based on an optimizing algorithm in
\cite{don05}: to have homogeneous degree, node distance,
betweenness, and loop distributions. But these rules are observed
from simulations,  theoretical proofs are not available by now. And,
sometimes, these rules are contrary to each other. For example,
comparing $C_6+e\{1,3\}$ with $C_6+e\{1,4\}$, one can find that the
cycle of $C_6+e\{1,4\}$ is more homogeneous, but the average node
distance of $C_6+e\{1,3\}$ is smaller. It seems that the importance
of these rules should be ordered. Although some rules are provided
in \cite{don05}, optimizing the network structure for better
synchronizability is still a hard problem, since it is possible that
the  optimizing algorithm converges to a suboptimal solution.

Other than the rules for optimization,  complementary graphs can be
used to characterize the  network synchronizability \cite{duan07}.
For a given graph $G$, the complementary graph  of $G$, denoted by
$G^c$, is the graph containing all the nodes of $G$ and all the
edges that are not in $G$.  For eigenvalues of graphs and
complementary graphs, the following lemma is useful (see
\cite{mer94, mer98} and references therein).

{\it Lemma 5}\,\, For any given graph $G$, the following statements
hold:

(i) \, $\lambda_N(G)$, the largest eigenvalue of $G$, satisfies
$\lambda_N(G)\leq N.$

(ii) \, $\lambda_N(G)= N$ if and only if $G^c$ is disconnected.

(iii)\, If $G^c$ is disconnected and  has (exactly) $q$ connected
components, then the multiplicity of $\lambda_N(G)=N$  is $q-1$.

(iv) \, $\lambda_i(G^c)+\lambda_{N-i+2}(G)=N,\quad 2\leq i\leq N$.

For example, the complementary graph of $C_{5o}$ is disconnected
(see Fig. 8) and the largest eigenvalue of $C_{5o}$ is 5, the number
of nodes. Its smallest nonzero eigenvalue $\lambda_2=2$ can be
easily obtained by computing the largest eigenvalue of its
complementary graph. Further, according to the complementary graph,
adding one more edge to graph $C_{5o}$ can not enhance its
synchronizability. However, if adding two more edges to $C_{5o}$,
e.g., $e\{1,5\}$ and $e\{3,5\}$, then the synchronizability
increases to $r=\frac{3}{5}$. The corresponding complementary graph
becomes the complementary graph of cycle $C_4$ (see Fig. 9). Cycle
$C_4$ and its complementary graph are very important in graph theory
\cite{mer98} (see the section below for their further applications).

For a given graph $G$, if its complementary graph is disconnected
and includes two separated graphs $G_1$ and $G_2$, then by Lemma 5
the synchronizability of $G$ is
$r(G)=\frac{N-\max\{\lambda_{max}(G_1), \lambda_{max}(G_2)\}}{N}$,
where $N$ is the number of nodes of $G$ and $\lambda_{max}$ denotes
the maximum eigenvalue of the corresponding Laplacian matrix. It is
well known that the complementary graphs of bipartite graphs are
disconnected \cite{duan07, pan02}, so the synchronizability of
bipartite graphs can be simply analyzed by the above method.
Actually, $C_{5o}$ in Fig. 6 is a bipartite graph. Obviously, better
understanding and careful manipulation of complementary graphs are
useful for enhancing the network synchronizability (see the section
below for further applications of complementary graphs).

\begin{center}
\unitlength=1cm \qquad \hbox{\hspace*{0.1cm}  \epsfxsize5cm
\epsfysize6cm \epsffile{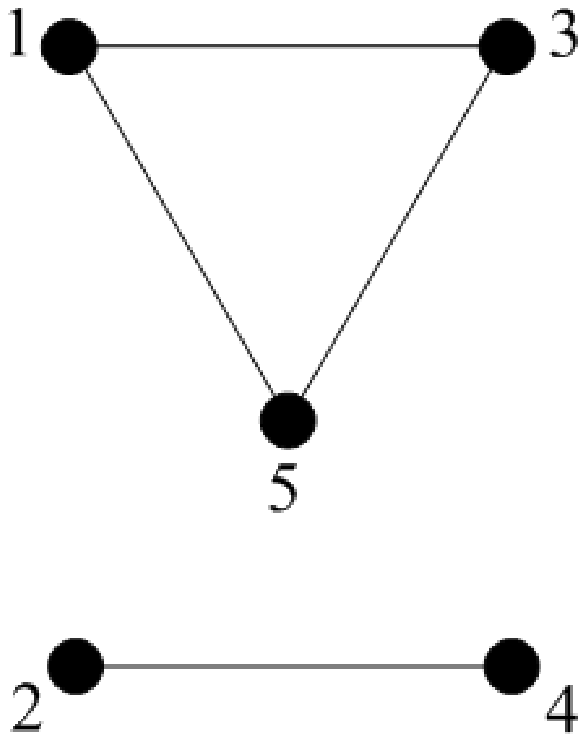}\qquad \epsfxsize3.5cm
\epsfysize5.5cm \epsffile{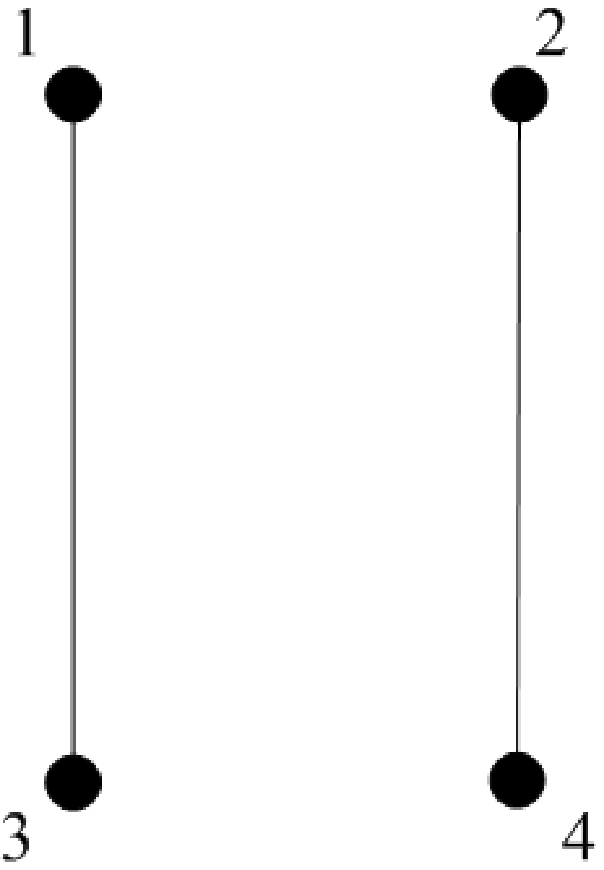} }
\end{center}
\vskip -0.3cm \centerline{ \qquad\qquad Fig. 8 \,\, Graph $C_{5o}^c$
\qquad\qquad \quad \quad Fig. 9 \,\, Graph $C_4^c$}

\section{Are networks with more edges easier to synchronize?}

For a given graph $G$, let ${\cal V}$ and ${\cal E}$ denote the sets
of nodes and edges of $G$, respectively.  A graph $G_1$ is called an
induced subgraph of $G$, if the node set ${\cal V}_1$ of $G_1$ is a
subset of ${\cal V}$ and the edges of $G_1$ are all edges  in ${\cal
E}$. In this section, subgraphs and complementary graphs are used to
discuss network synchronizability.

In the concern of  optimizing  network structures, an interesting
question is whether networks with more edges are easier to
synchronize. In order to answer this question, the following lemma
is needed.

{\it Lemma 6}\,\, For any given graph $G$, suppose $G_1$ is its
induced subgraph including all nodes of $G$ with the maximum degree
$d_{max}$. If $G_1$ includes a cycle $C_{2k}$ with even nodes $2k
(k\geq 2)$ as an induced subgraph, then the largest eigenvalue of
$G$ satisfies $\lambda_N (G)\geq d_{max}+2.$

{\it Proof}\,\, By Lemma 3, for any cycle $C_{2k}$ with even nodes,
its largest eigenvalue is 4. And since the degree of every its node
is 2, $-2$ must be an eigenvalue of the adjacency matrix of
$C_{2k}$. Let $L_1$ be the sub-matrix of the Laplacian matrix of $G$
associated with all the nodes in $G_1$.  By the assumption, one has
$$(d_{max}+2)I-L_1=2I+A(G_1)\not > 0,$$
where $A(G_1)$ is the adjacency matrix of $G_1$. This implies that
the largest eigenvalue of $G_1$ is larger than or equal to
$d_{max}+2$. Thus, Lemma 1 leads to the result directly. \hfill
$\Box$

{\it Remark 1}\,\, Besides Lemma 2, there are few results on lower
bounds of the largest eigenvalue of Laplacian matrices in graph
theory \cite{mer94, pan02}. Since networks with good
synchronizability always have homogeneous degree distributions,
Lemma 6 is very useful for the study of network synchronization.




{\it Theorem 2}\,\, For any graph $G$ with 16 edges on 10 nodes, its
eigenratio  is bounded by  $r(G)< \frac{2}{5}$.

{\it Proof}\,\, If the largest node degree of $G$ is $d_{max}\geq
6$, then the smallest node degree must stisfy $d_{min}\leq 2$. The
conclusion follows directly from Lemma 2. In order to have good
synchronizability, the degree distribution of $G$ should be
homogeneous. Then, first suppose that $G$ has 8 nodes with degree 3
and two nodes with degree 4. In this case, by Lemma 2, the largest
eigenvalue of $G$ is $\lambda_{10} (G)> 5$.

In what follows, consider the largest eigenvalue of the
complementary graph $G^c$. By the above discussion,  $G^c$ must have
 8 nodes with degree 6 and two nodes with degree 5. Suppose $G_1$
is the subgraph of $G^c$ composing of 8 degree-6 nodes.  By direct
computing, $G_1$ must have 19 or 20 edges, and the degree of every
its node is at least 4. Hence, $G_1^c$ has 9 or 8 edges and the
degree of every its node is at most 3. If the largest eigenvalue of
$G_1$ is 8, i.e., $G_1^c$ is disconnected (Lemma 5), then the
largest eigenvalue of $G^c$ is larger than or equal to 8. Therefore,
the smallest nonzero eigenvalue of $G$ is $\lambda_2(G) \leq
10-8=2$. By the above discussion, the theorem obviously holds.
Hence, suppose both $G_1$ and $G_1^c$ are connected. Then, $G_1$
must have a cycle $C_4$ as an induced subgraph. This holds if and
only if $G_1^c$ has $C_4^c$ (see Fig. 9) as an induced subgraph.
With only 9 or 8 edges having node degree at most 3, drawing $G_1^c$
directly one can easily reach the conclusion. By Lemma 6, the
largest eigenvalue of $G^c$ must be larger than or equal to 8.
Repeating the above discussion concludes the proof.

When $G$ has 9 nodes with degree 3 and one node with degree 5, the
proof can be similarly completed. \hfill $\Box$

{\it Remark 2}\,\, Theorem 2 shows that there is not  a graph $G$
with 16 edges on 10 nodes whose synchronizability is $r(G) \geq
\frac{2}{5}.$ However, there does exist a graph $\Gamma_1$ with 15
edges on 10 nodes whose synchronizability  is
$r(\Gamma_1)=\frac{2}{5}$ (see Fig. 10),  consistent with the result
of \cite{don05}. This clearly shows that networks with more edges
are not necessarily easier to synchronize. In fact, by the optimal
result of \cite{don05}, $r=\frac{2}{5}$ is the optimal
synchronizability for graphs with 15 edges on 10 nodes. For any
graph $G$ with 16 edges on 10 nodes, if both $G$ and $G^c$ have
cycles with even nodes, then by Lemma 6 and Theorem 2, $r(G)\leq
\frac{2}{6}=\frac{1}{3}$. Therefore, adding one more edge definitely
decreases the synchronizability. The existence of cycles with even
nodes can be easily tested by drawing graphs, so Lemma 6 is very
useful for analyzing the synchronizability of homogeneous networks.
Actually, the graph shown in Fig. 10 is quite homogeneous in
structure \cite{don05}. With one more edge being added, such a
structure is destroyed. It is therefore easy to understand why
adding more edges do not necessarily result in better
synchronizability.

\begin{center}
\unitlength=1cm \qquad \hbox{\hspace*{0.1cm} \epsfxsize7cm
\epsfysize7cm \epsffile{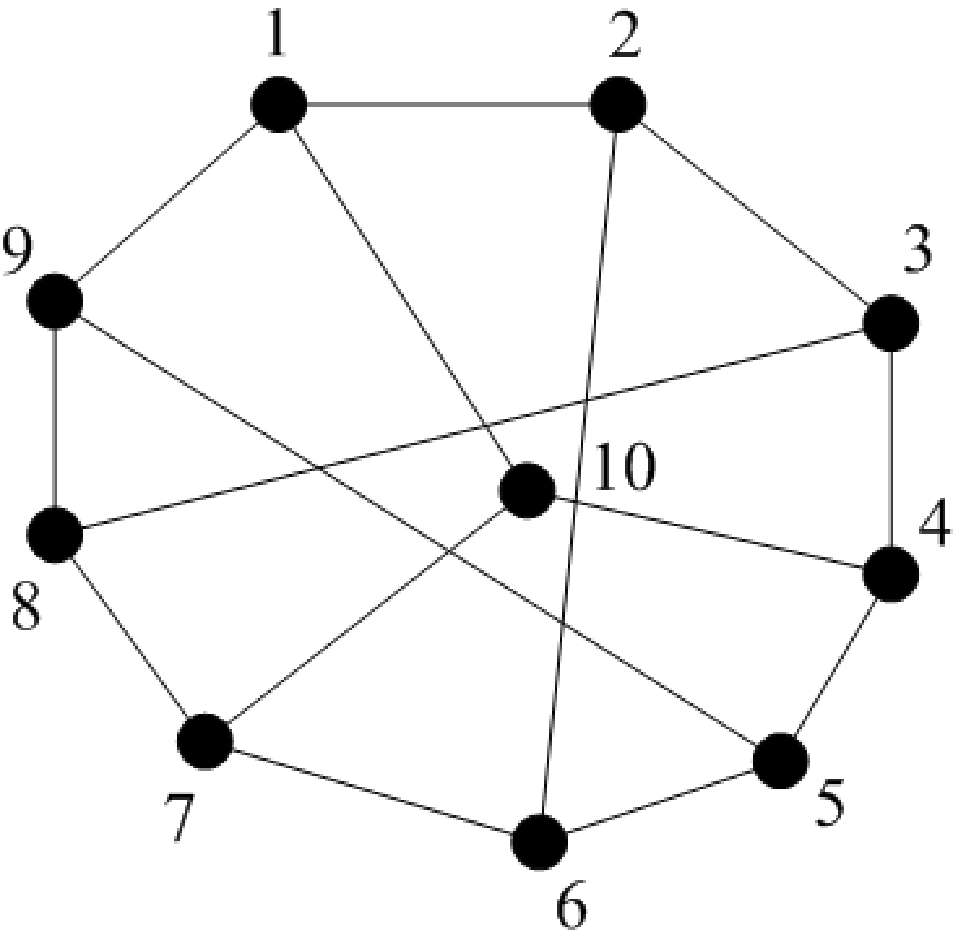} }
\end{center}
\vskip -0.3cm \centerline{  Fig. 10 \,\, Graph $\Gamma_1$,
$r(\Gamma_1)=\frac{2}{5}$}

{\it Remark 3}\,\,Fig. 11 shows a new graph $\Gamma_2$ with 20 edges
on 10 nodes. It also has quite homogeneous  structural
characteristics as discussed in \cite{don05}.  In fact, the
betweenness centrality of each node of $\Gamma_1$ is 6, larger than
that of $\Gamma_2$, 5. But, the synchronizability of graph
$\Gamma_2$ is worse than that of graph $\Gamma_1$, contrary to the
result of \cite{hong04}. So far, the existing theories \cite{nis03,
hong04, don05, zhao06} can not explain why the synchronizability of
$\Gamma_1$ is better than that of $\Gamma_2$. This shows the
complexity of the relationships between the synchronizability and
network structural characteristics. Although $\Gamma_2$ has the
property of  homogeneity, another question is whether there exists
another graph with 20 edges on 10 nodes having better
synchronizability than that of $\Gamma_1$ or $\Gamma_2$?  If the
answer is negative, it implies that generally there are many
redundant edges in a network with respect to  its synchronizability.
This kind of questions are still open today.

\begin{center}
\unitlength=1cm \qquad \hbox{\hspace*{0.1cm} \epsfxsize7cm
\epsfysize7cm \epsffile{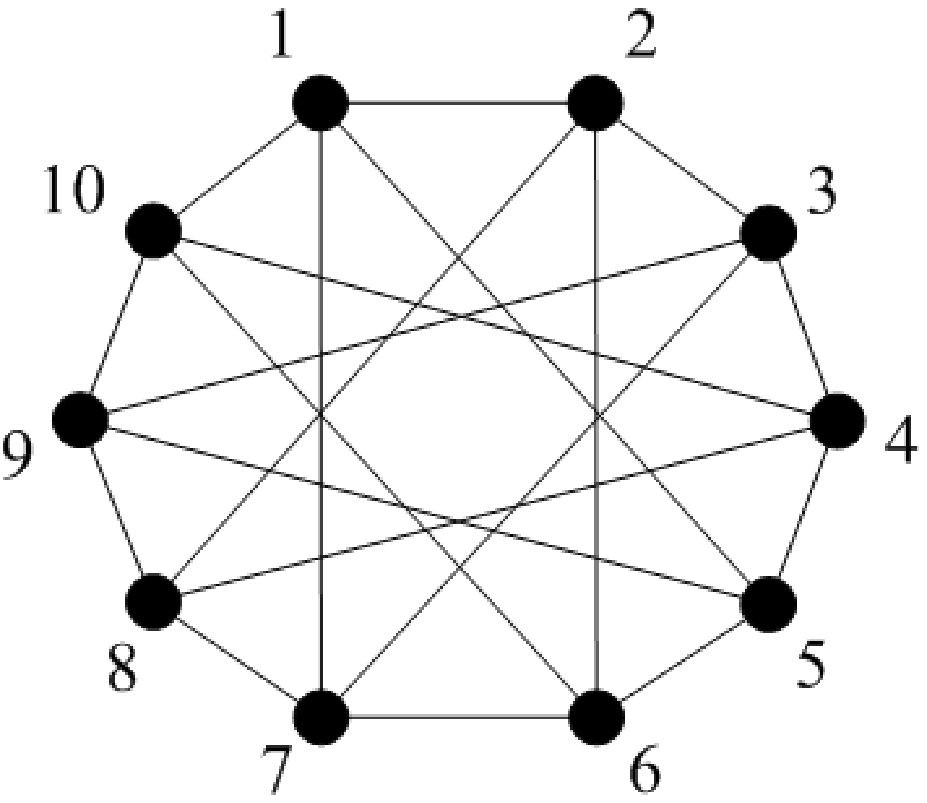} }
\end{center}
\vskip -0.3cm \centerline{  Fig. 11 \,\, Graph $\Gamma_2$,
$r(\Gamma_2)=\frac{2.7639}{7.2361}\approx
0.382<r(\Gamma_1)=\frac{2}{5}$}

\section{Some examples}

In this section, some examples are given to show the changes of the
synchronizability versus the addition of edges.

{\it Example 1}\,\, The synchronizability changes by adding edges to
graphs with cycles are shown in Figs. 12 and 13, where their initial
graphs  are $C_{10}$ and $C_{50}$, respectively, and $m_{add}$
denotes the number of added edges. The figures in (a)s show the
synchronizability changes during the process of adding edges with
degree homogeneity (i.e., guaranteeing the node degrees be as
homogeneous as possible during edge-adding). The figures in (b)s
show the cases corresponding to random edge-adding. Naturally, the
corresponding synchronizabilities in (a)s are better than those in
(b)s, since degree homogeneity is an important property for networks
to achieve good synchronizabilities. In all graphs, it is shown that
the synchronizability globally increases but locally fluctuates.
According to Theorem 2 and Remark 2, this is the expected
phenomenon.

\begin{center}
\unitlength=1cm \qquad \hbox{\hspace*{0.1cm}  \epsfxsize7cm
\epsfysize7cm \epsffile{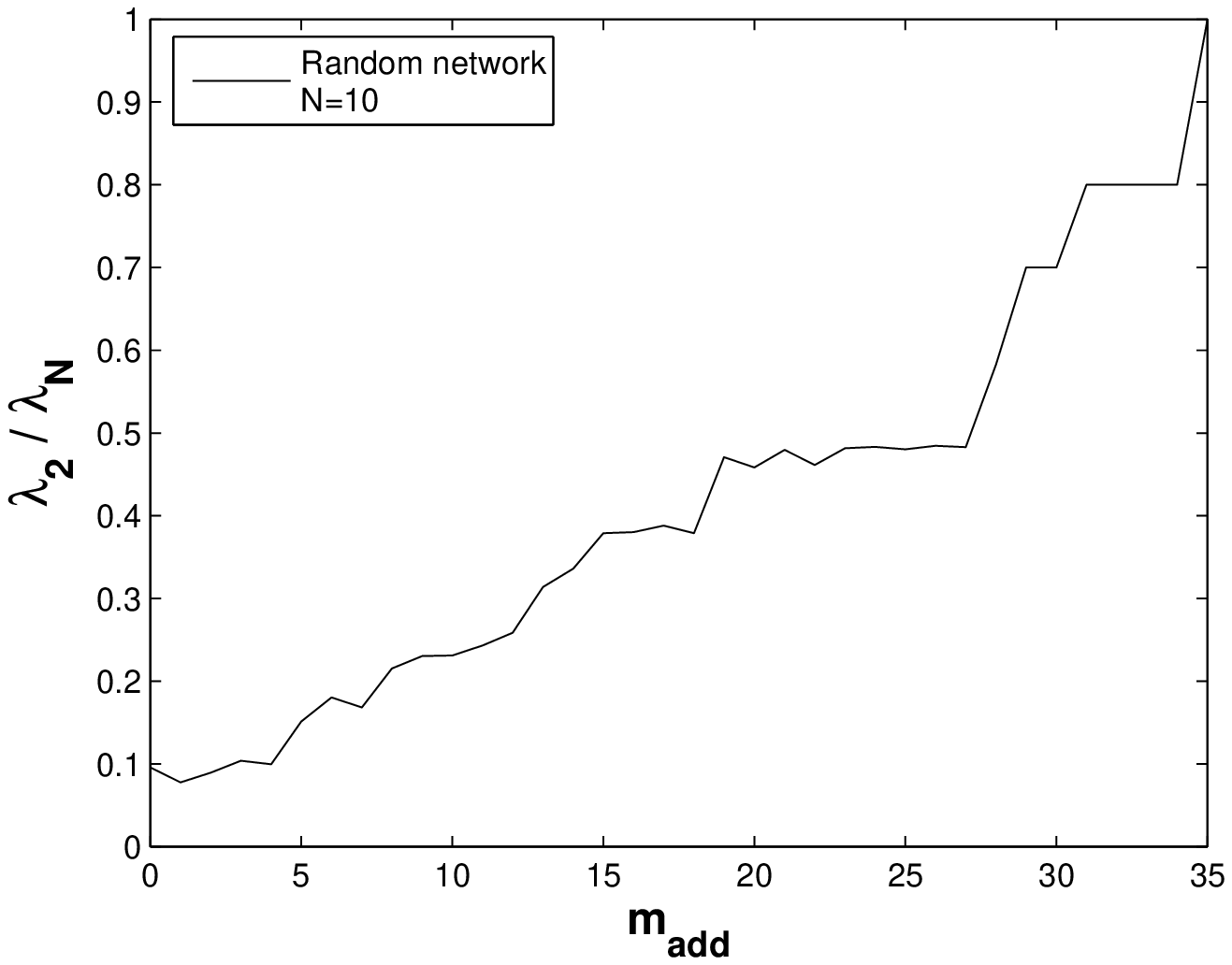}\quad \qquad\qquad\epsfxsize7cm
\epsfysize7cm \epsffile{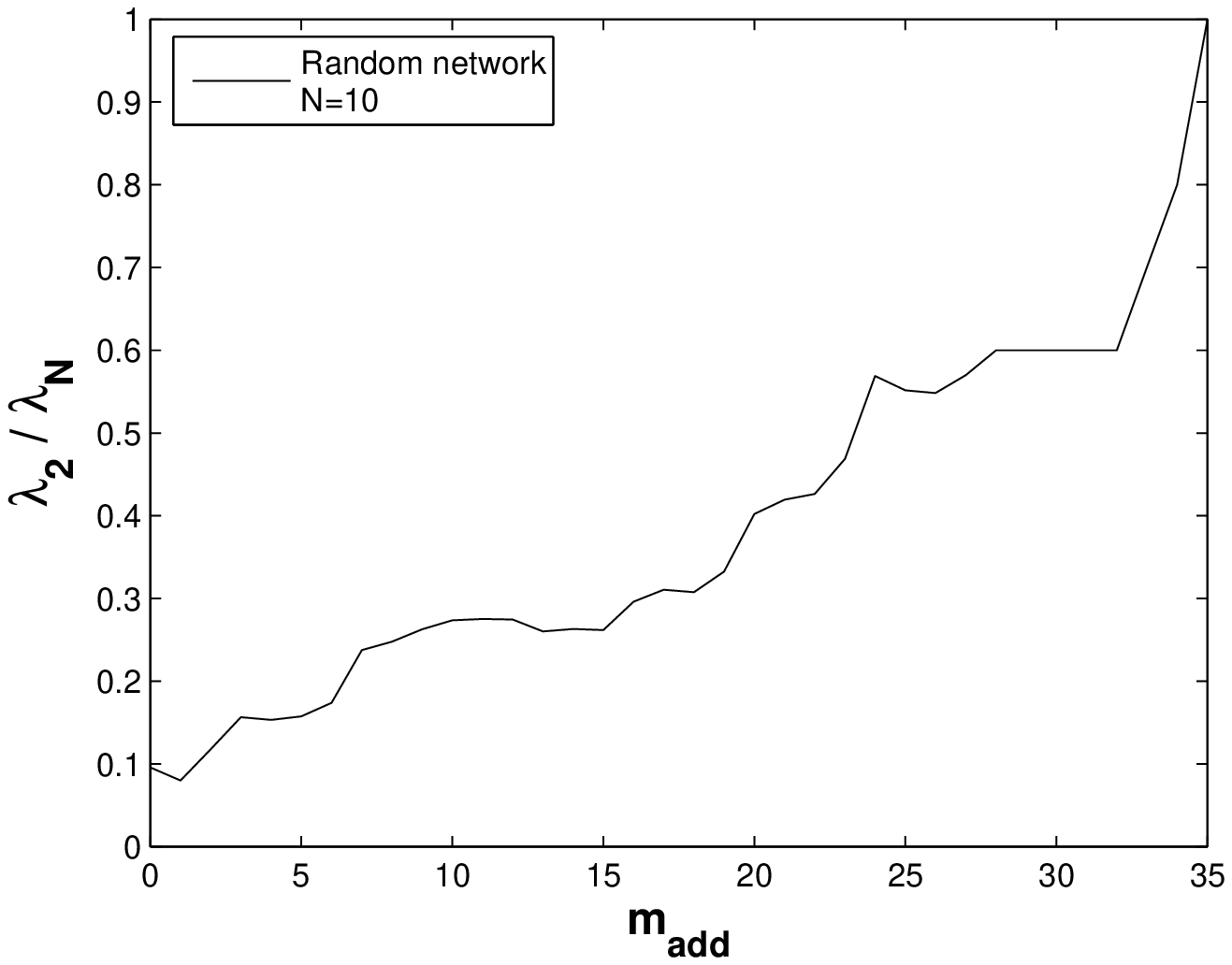} }
\end{center}
\vskip -0.3cm \centerline{(a) \,\, Adding edges with degree
homogeneity  \quad\qquad (b) \,\, Randomly adding edges}
\centerline{ Fig. 12 \,\, The synchronizability changes of graphs
obtained from $C_{10}$ by adding edges }

\begin{center}
\unitlength=1cm \qquad \hbox{\hspace*{0.1cm}  \epsfxsize7cm
\epsfysize7cm \epsffile{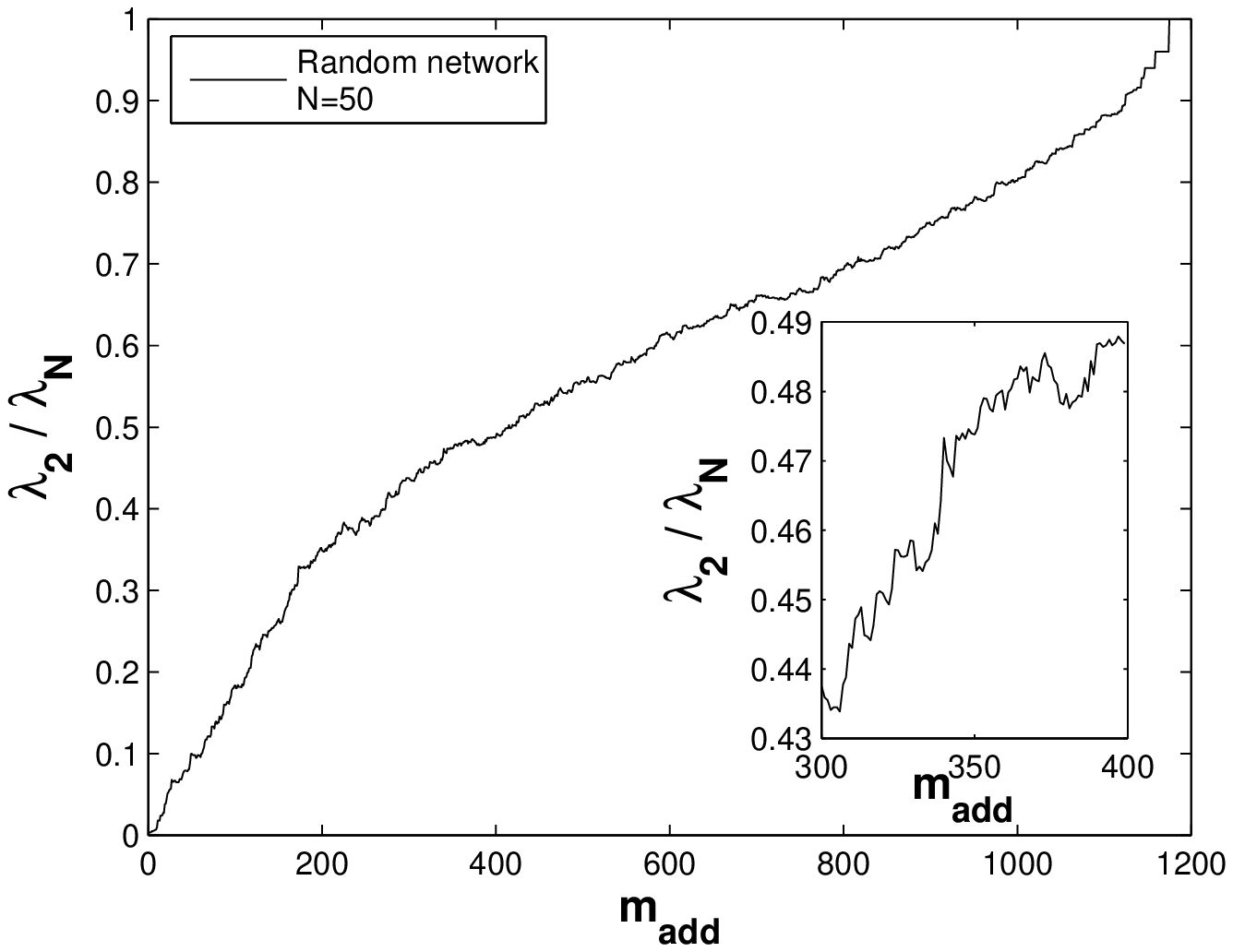} \qquad\epsfxsize7cm
\epsfysize7cm \epsffile{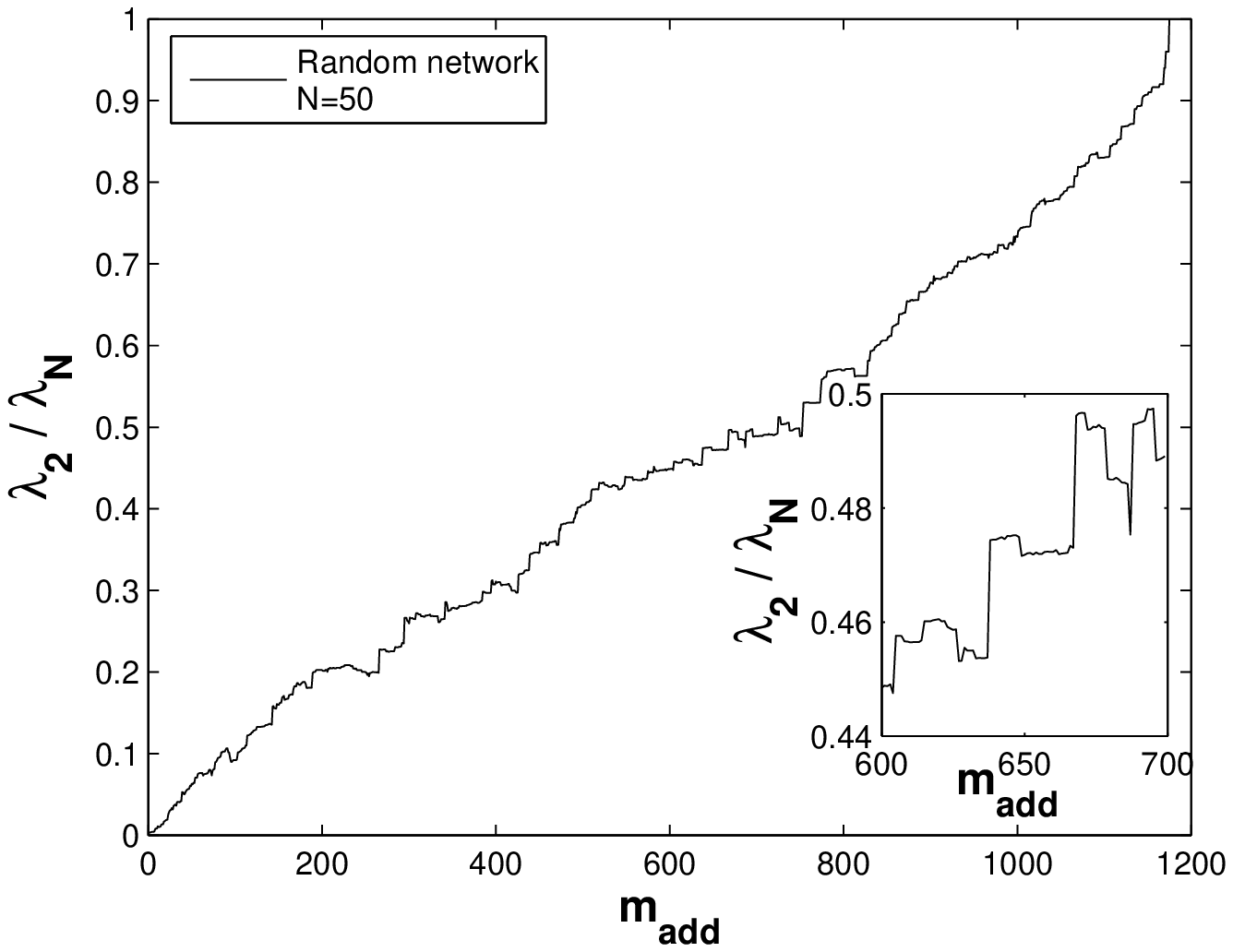} }
\end{center}
\vskip -0.3cm \centerline{(a) \,\, Adding edges with degree
homogeneity  \qquad\quad (b) \,\, Randomly adding edges}
\centerline{ Fig. 13 \,\, The synchronizability changes of graphs
obtained from $C_{50}$ by adding edges }

{\it Example 2}\,\,The synchronizability changes of graphs obtained
from a scale-free graph by randomly adding edges are shown in Fig.
14, for which, the same conclusion can be drawn as in Example 1.

\begin{center}
\unitlength=1cm \qquad \hbox{\hspace*{0.1cm}  \epsfxsize8cm
\epsfysize8cm \epsffile{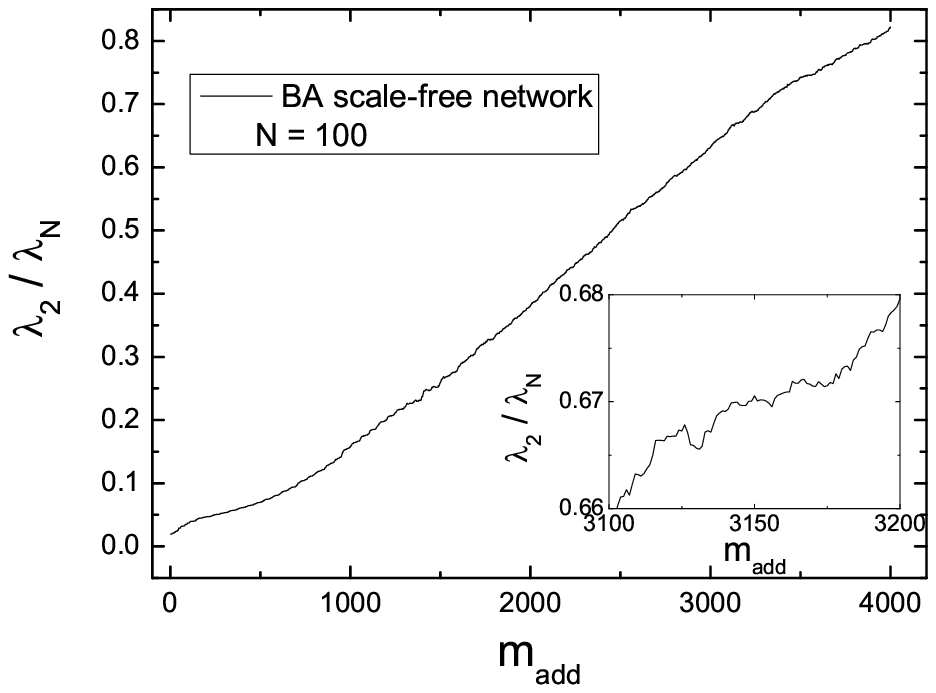}\epsfxsize8cm \epsfysize8cm
\epsffile{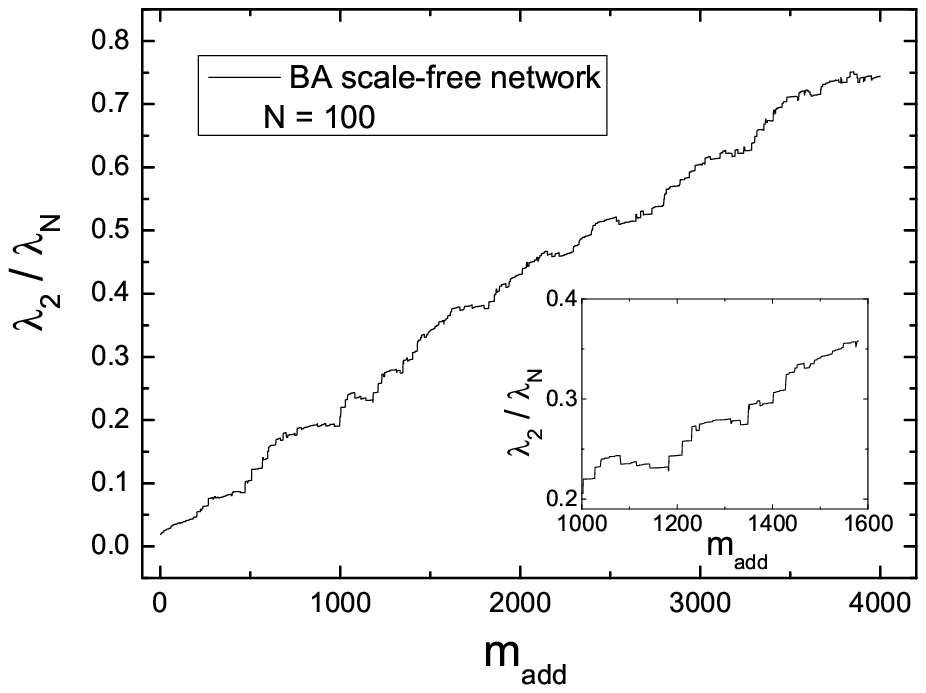} }
\end{center}
\vskip -0.3cm \centerline{(a) \,\, Adding edges with degree
homogeneity  \qquad\quad (b) \,\, Randomly adding edges}
\centerline{ Fig. 14 \,\, The synchronizability changes of graphs
obtained from } \centerline{ a scale-free graph by adding edges}

\section{Conclusion}

In this paper, the relationship between the network
synchronizability and the edge distribution of the associated graphs
has been studied. It has been proved that the synchronizability
definitely decreases if one edge is being added to a cycle with $N$
($N\geq 5$) nodes. However, it has also been shown that the
synchronizability can be improved by changing the network structure.
Further, some examples have shown that some networks with more
edges, unexpectedly, have worse synchronizabilities even if the
network structures are in some sense optimized. This implies that,
for network synchronization, generally there are redundant edges,
which do not make any contribution to synchronization but may
actually destroy the synchronizability. In addition, an example of a
graph with 20 edges on 10 nodes has been provided to show that the
existing theories can not explain why it has worse synchronizability
than that of a graph with 15 edges on 10 nodes. Some other examples
have also been given to show that the network synchronizability
globally increases but locally fluctuates due to edge-adding.
According to these results, in practical synchronization problems,
the synchronizability and the number of communication edges should
have a coordinative relation. And  one may utilize the redundant
edges to improve robustness or other network properties. These kinds
of important questions remain open for further research  in the
future.

\hspace*{34pt}

\end{document}